\documentclass[doublecol]{epl2} 
\usepackage{graphicx}  
\usepackage{dcolumn}   
\usepackage{bm}        
\usepackage{amssymb}   
\usepackage{amsmath}
\usepackage{braket}
\usepackage[mathscr]{euscript}
\usepackage{color}
\usepackage{epsfig,color}
\usepackage{hyperref}

\DeclareMathOperator{\sech}{sech}

\title{Nonlinear compression of temporal solitons in an optical waveguide via inverse engineering}

\author{Koushik Paul \inst{1} \and Amarendra K. Sarma\inst{1}}

\institute{                   
  \inst{1} Department of Physics, Indian Institute of Technology Guwahati, Guwahati-781039, Assam, India
}

\abstract{We propose a novel method based on the so-called shortcut to adiabatic passage techniques to achieve fast compression of temporal solitons in a nonlinear waveguide. We demonstrate that soliton compression could be achieved, in principle, at an arbitrarily small distance by inverse engineering the pulse width and the nonlinearity of the medium. The proposed scheme could possibly be exploited for various short-distance communication protocols and may be even in nonlinear guided wave-optics devices and generation of ultrashort soliton pulses.}

\begin{document}

\maketitle

\section{Introduction}
Solitons are self-reinforcing nonlinear localized waves that propagate without spreading and have particle-like properties \cite{b.b1,b.b2}. Solitons are universal in nature and observed in various platforms ranging from plasmas and fluids to chemical, biological, solid-state, optical and magnetic systems \cite{b.b2,b.b3,b.p4,b.p5,b.p6}, and even to elementary particles \cite{b.p7}. In optics, temporal bright solitons are of particular interest owing to its various possible applications, specifically for long distance optical communications and all-optical controllability of pulse shape \cite{b.b8}. Temporal bright solitons are formed due to precise interaction of linear group velocity dispersion (GVD) and positive self-phase modulation  (SPM). Recently, temporal soliton pulse compression studies are getting tremendous boost after a couple of experimental studies. For example, Gerome {\it et al.} reported soliton compression in a tapered hollow-core photonic bandgap fiber \cite{b.p9}. Colman {\it et al.} demonstrated pulse compression based on high-order solitons in photonic crystal waveguides \cite{b.p10}. Blanco-Redondo  {\it et al.}, experimentally demonstrated soliton-effect pulse compression of picosecond pulses in silicon \cite{b.p11}. 
In another relatively recent development, a wide class of solitonlike self-similar waves called optical similaritons are discovered \cite{b.p12}. Unlike solitons, which are normally described by the so-called nonlinear Schrodinger equation with constant GVD and SPM parameters in a passive waveguide, similaritons correspond to situation where both the GVD and the SPM parameter changes along the length of the waveguide with distributed loss or gain \cite{b.b8,b.p12,b.p13,b.p14,b.p15,b.p16}. Similariton pulse compression is also extensively studied by various authors owing to its practical significance in amplifying medium \cite{b.p17,b.p18, b.p19, b.p20, b.p21, b.p22, b.p23}.
 In this work, we focus on a passive nonlinear Schrodinger system with constant GVD parameter and distributed nonlinear parameter. We consider a passive waveguide with varying nonlinearity and study the compression of temporal solitary waves with an engineered nonlinearity. We propose a novel method based on the so-called shortcut to adiabatic passage techniques to achieve fast compression of solitary waves. Shortcut to adiabatic (STA) passage methods, in particular the transitionless quantum driving (TQD) algorithm \cite{b.p24,b.p25} and Lewis-Reisenfeld invariant (LRI) \cite{b.p26,b.p27} based approach has gained immense popularity in recent years in diverse areas of physics \cite{b.p28}.  STA methods, developed originally in the context of quantum mechanics, now find its place in various theoretical proposals and even in experiments \cite{b.p28.1,b.p28.2}. It has been explored in speeding up, dynamics in spinor condensates \cite{b.p29}, quantum annealing \cite{b.p30}, power switching in a waveguide coupler \cite{b.p30.1, b.p30.2, b.p30.3,b.p31}, engineering non-hermitian systems \cite{b.p32}, atom cooling in harmonic traps \cite{b.p33}, entanglement preparation \cite{b.p34} and so on. In the context of soliton compression, adiabatic methods have been explored earlier in various contexts \cite{b.p35,b.p36,b.p37}. However, very few attempts have been made to achieve soliton compression using STA methods. In this regard, one particular study worth mentioning is that of Jing Li {\it et al.}, where they proposed a method to achieve controlled compression of soliton matter waves in harmonic traps by tunable interaction using STA \cite{b.p37.1}. In context of fiber optics, Mamychev {\it et al.} experimentally demonstrated the adiabatic soliton compression in an optical fiber as a result of the combined effects of delayed nonlinear response and higher-order dispersion \cite{b.p36}. Quiroga-Teixeiro {\it et al.} proposed a scheme for efficient soliton compression in a monomode optical fiber by fast adiabatic amplification \cite{b.p37}. The primary disadvantage of such schemes is that one needs a very long fiber. In this work, for the first time to the best our knowledge, we propose a scheme to obtain temporal soliton compression in a nonlinear waveguide at an arbitrarily small length using the shortcut to adiabatic passage technique. The scheme proposed in this work is a combination of the Lagrangian variational approach with STA, first introduced by Jing Li {\it et al.} \cite{b.p37.1} in the context of soliton matter waves. It is worthwhile to note that the soliton compression was first elaborated theoretically, using the variational approximation, by Anderson {\it et al.} \cite{b.p37.2}. In this work, authors suggested a scheme to compress soliton by engineering the dispersion but keeping nonlinear coefficient constant. Incidentally, this theoretical prediction was realised experimentally by K. Bertilsson {\it et al.} \cite{b.p37.3}.
 
 The article is organised as follows: the next section describes the theoretical model. The third section discusses compression of temporal solitons via adiabatic methods. The disadvantage of the method is also briefly discussed. Finally, the fourth section discusses compression of temporal solitons using the inverse engineering technique followed by conclusions in the last section.

\section{The model}
The general nonlinear wave equation governing pulse propagation in an inhomogeneous passive nonlinear waveguide can be written as \cite{b.p12}:
\begin{equation}
i\frac{\partial \psi}{\partial z} + \frac{\beta_2(z)}{2}\frac{\partial^2\psi}{\partial t^2} + \gamma(z) |\psi|^2\psi = 0,
\label{WaveEq}
\end{equation}
where, $\psi$ is the beam envelope, $\beta_2$ the group velocity dispersion (GVD) parameter and  $\gamma$ is the nonlinear parameter. Eq. (\ref{WaveEq}) could be written in dimensionless form as follows:
\begin{equation}
i\frac{\partial u}{\partial \xi} + \frac{1}{2}\frac{\partial^2u}{\partial \eta^2} +\gamma|u|^2 u = 0,
\label{NLSE}
\end{equation}
where $u(\xi,\eta)$ is the normalized amplitude and 
\begin{equation}
\psi = \sqrt{P_0}Nu, \qquad \eta= t/T_0, \qquad \xi = z/L_D
\label{para}
\end{equation}
in which $\xi$  and  $\eta$ are the normalized propagation distance and time respectively, $P_0$ is the peak power of the incident pulse and $T_0$ is the initial pulse width. Also $L_D = T_0^2/|\beta_2|$ is the dispersion length and $N = \sqrt{\gamma P_0 L_D}$ is the so called soliton order. In this work, we assume that $\beta_2$ is a constant and independent of the propagation distance, while $\gamma$ has a specific $z$ dependence. It is worth mentioning that, Eq.(\ref{WaveEq}) is valid for input pulse as short as $T_0=5 ps $ . For $T_0< 5 ps $, it is necessary to include higher order nonlinear and dispersive effects such as third-order dispersion, self-steepening and intrapulse Raman scattering effects \cite{b.b8}.

Eq. (\ref{NLSE}) has the following well-known (scaling) bright solitary wave solution, given by \cite{b.b38}:
\begin{equation}
u(\xi,\eta) = A(\xi)\sech \bigg[ \frac{\eta'}{\alpha(\xi)}\bigg]\exp\big[ i\beta(\xi)\eta'^2 +i\epsilon(\xi)\eta'+i\phi(\xi) \big]
\label{Solitary}
\end{equation}
where $\eta'=\eta-\zeta(\xi)$. Here $A(\xi)$, $\alpha(\xi)$, $\beta(\xi)$, $\epsilon(\xi)$, $\zeta(\xi)$ and $\phi(\xi)$ represent the amplitute, width, chirp, velocity, center position and phase respectively. These are all real functions.  It is easy to get that, $\int_{-\infty}^{\infty} |u|^2 \,d\eta = 2\alpha A^2 = 2N$,where $N$ is the normalization parameters, such that, $A = \sqrt{N/\alpha}$.

The Lagrangian density \cite{b.b38.1} corresponding to Eq.~(\ref{NLSE}) is:
\begin{equation}
\mathscr{L} = \frac{1}{2}\bigg(\frac{\partial u}{\partial \xi} u^* - \frac{\partial u^*}{\partial \xi} u\bigg) - \frac{1}{2} \bigg\lvert \frac{\partial u}{\partial \eta} \bigg\rvert^2 + \frac{1}{2} \gamma(\xi)|u|^4
\label{Lagrange}
\end{equation}
Now, inserting Eq.~(\ref{Solitary}) into Eq.~(\ref{Lagrange}), we can find the Lagrangian, $L = \int_{-\infty}^{\infty}\mathscr{L} \,d\eta$. Using the Euler-Lagrange formulas, $\frac{\delta L}{\delta q} = 0$ , where $q$ represents one of the parameters $\alpha(\xi)$, $\beta(\xi)$, $\epsilon(\xi)$ and $\zeta(\xi)$, we obtain the following set of coupled differential equations:
\begin{subequations}
\begin{align}
\label{coup0}
\frac{\,d\alpha}{\,d\xi} &= 2\alpha\beta\\
\label{coup1}
\frac{\,d\beta}{\,d\xi} &= \frac{4}{\pi^2\alpha^4} - 2\beta^2 - \frac{2\gamma N}{\pi^2 \alpha^3} +f_0 \\
\label{coup2}
\frac{\,d\epsilon}{\,d\xi} &= 0\\
\label{coup3}
\frac{\,d \zeta}{\,d\xi} &= \epsilon
\end{align}
\label{amps}
\end{subequations}
It should be noted that $\frac{\delta L}{\delta \phi} = 0$ , implying that $\phi$ does not play any role in the variational dynamics and hence we put $\phi=0$ in the rest of the work.
Above set of equations could be simplified to the following equations involving only the two main parameters, $\alpha$ and $\zeta$:
\begin{eqnarray}
\label{alpha}
\ddot{\alpha} &=& \frac{8}{\pi^2\alpha^3} - \frac{4\gamma N}{\pi^2 \alpha^2}\\
\label{zeta}
\ddot{\zeta} &=& 0
\end{eqnarray}
where dot refers to derivative with respect to $\xi$. It is obvious from the above equation that the width of the solitary wave, $\alpha$, is dependent of the Kerr parameter $\gamma$. The realistic solution of Eq.~(\ref{zeta}) is simply $\zeta(\xi)=0$. Thus, we will focus on Eq.~(\ref{alpha}) only for inverse engineering.

\section{Compression via adiabatic process}

Our objective is to achieve fast and perfect solitary wave compression, by judicious designing of the nonlinear Kerr parameter $\gamma$, from initial state $u(0,\eta)$ to the final state $u(\xi_f,\eta)$. The expression for $u(0,\eta)$ and $u(\xi_f,\eta)$ are given as follows:
\begin{subequations}
\begin{align}
\label{U0}
u(0,\eta) &= \sqrt{\frac{N}{\alpha(0)}}\sech\bigg[\frac{\eta}{\alpha(0)}\bigg]\exp\bigg(i\frac{\dot{\alpha}(0)}{2\alpha(0)}\eta^2 \bigg)\\
\label{Uf}
u(\xi_f,\eta) &= \sqrt{\frac{N}{\alpha(\xi_f)}}\sech\bigg[\frac{\eta}{\alpha(\xi_f)}\bigg]\exp\bigg(i\frac{\dot{\alpha}(\xi_f)}{2\alpha(\xi_f)}\eta^2 \bigg)
\end{align}
\end{subequations}
It may be useful to note that Eq.~(\ref{alpha}) is analogous to a classical particle moving kinetic energy $\frac{1}{2}\dot{\alpha}^2$, and potential energy of the following form:
\begin{equation}
V(\xi) = \frac{4}{\pi^2\alpha^2} - \frac{4\gamma N}{\pi^2 \alpha}
\label{potential}
\end{equation}
This enables us to get the adiabatic reference for soliton compression. $\partial V/\partial \alpha = 0$ gives the minimum point of potential, {\it i.e.},
\begin{equation}
\alpha_m(\xi) = \frac{2}{\gamma(\xi)N}
\label{MinPot}
\end{equation}
Eq.~(\ref{MinPot}) yields the minimum perturbed potential of an effective particle, also giving us the adiabatic reference when Kerr nonlinearity $\gamma(\xi)$ is given. In this work, we choose the following switching function $\gamma(\xi)$ as given below: 
\begin{equation}
\gamma(\xi) = \gamma_0 + \lambda \big[ 1 +\tanh \big \{ \delta(\xi-\frac{\xi_f}{2})\big \} \big]
\label{NL}
\end{equation}
Here $\lambda$ and $\delta$ are the control parameters and $\xi_f$ is the final distance. $\gamma_0$ is the nonlinear parameter without the control. To understand the adiabatic reference one needs to solve Eq.~(\ref{alpha}) numerically and compare it with $\alpha_m(\xi)$ . It is found that $\alpha(\xi)$, the solution of Eq.~(\ref{alpha}) coincides with the minimum value of $\alpha_m(\xi)$ adiabatically.
\begin{figure}
\onefigure[height=6cm,width=8cm]{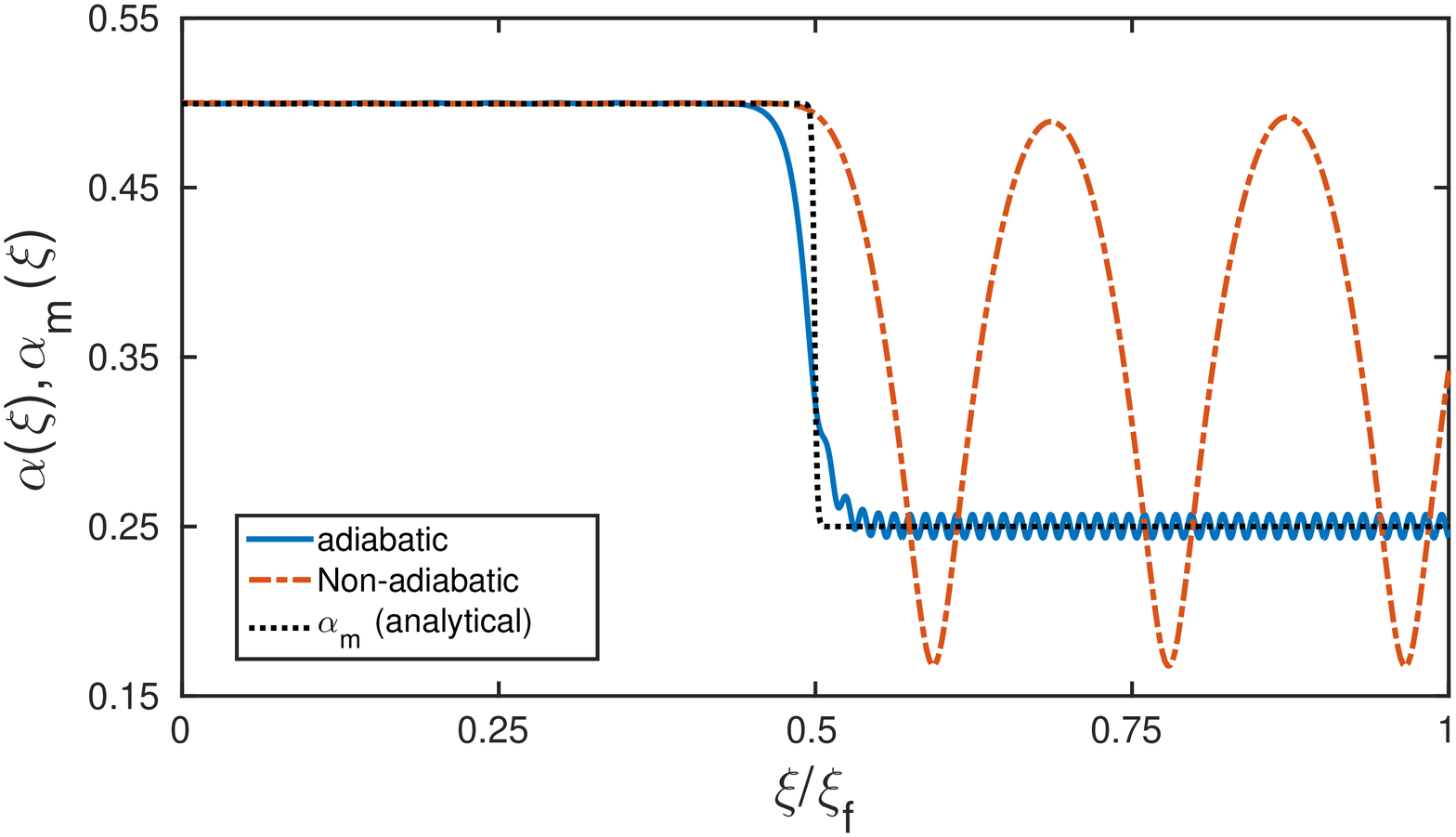}
\caption{(Color online) Comparison of the compression of Soliton width $\alpha(\xi)$ as a function of $\xi$ with $\gamma_0 = 2$. Adiabatic compression (solid blue) is achieved for $\lambda = 1$, $\delta = 1$, $\xi_f = 50$. Non adiabatic compression (dash-dotted brown) is shown for $\lambda = 1$, $\delta = 10$, $\xi_f = 5$. $\alpha_m(\xi)$ using Eq.~(\ref{MinPot}) (dotted black) follows adiabatic path exactly.}
\label{Fig1}
\end{figure}
Fig.~(\ref{Fig1}) demonstrates the evolution of temporal soliton width as a function of distance for both adiabatic and non-adiabatic cases. Here we compare the exact result obtained by solving Eq.~(\ref{alpha}) with the adiabatic reference, given by Eq.~(\ref{MinPot}). We observe that the exact result nearly matches the adiabatic result; the pulse gets compressed from its initial value
$\alpha=0.5$ to  $\alpha=0.25$ after propagating a distance, $\xi = 50$, for the chosen parameters. On the other hand, while propagating a distance on the order of $\xi = 5$ , the adiabatic reference is no longer followed and one cannot obtain effective pulse compression at such shorter distance. 
The compression of soliton width requires large propagation distance and low switching rate ( small $\delta$). It is quite difficult to achieve  same amount of compression in smaller propagation distances in such set up. However one can use inverse engineering approach in order to create desired compression within small $\xi_f$ value. It should be noted that amount of compression could be controlled by controlling $\lambda$. Higher compression ($\alpha(\xi_f) < 0.25$) can be achieved for larger $\lambda$ values.
\section{Compression via inverse engineering}

The inverse engineering technique based on fixed boundary conditions and ansatz is pioneered by Chen and Muga \cite{b.p27,b.p33,b.p39,b.p40}. To inverse engineer our system, we choose a set of predefined initial and final state of $\alpha(\xi)$ which coincides with $\alpha_m(\xi)$ and design it using polynomial ansatz \cite{b.p33,b.p39,b.p40} given by 
\begin{figure}
\onefigure[height=10cm,width=8cm]{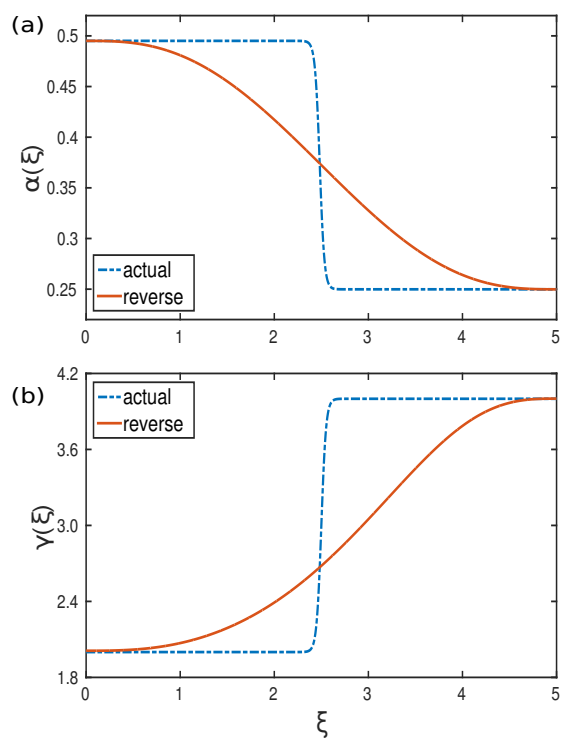}
\caption{(Color online) Controlling Soliton width using reverse engineering, (a) $\alpha(\xi)$ analytical (dash-dotted blue) and reverse engineered (solid brown) and (b) Non-linearity parameter $\gamma(\xi)$ as chosen in Eq.~(\ref{NL}) (dash-dotted blue) and reverse engineered (solid brown) with the parameters  $\lambda = 1$, $\delta = 10$, $\xi_f = 5$ and $\gamma_0 = 2$.}
\label{Fig2}
\end{figure}
\begin{equation}
\alpha(\xi) = \sum_{j=0}^{5} a_j \xi^j
\label{poly}
\end{equation}
However this needs the application of appropriate boundary conditions. To fix the initial and the final value of the width we set 
\begin{equation}
\alpha(0) = \alpha_m(0), \qquad \alpha(\xi_f) = \alpha_m(\xi_f)
\label{bound1}
\end{equation}
Another set of boundary conditions is required to satisfy the continuity of $\alpha(\xi)$, which are given by
\begin{eqnarray}
\label{bound2}
\dot{\alpha}(0) = \dot{\alpha}_m(0), \qquad \dot{\alpha}(\xi_f) = \dot{\alpha}_m(\xi_f) \\
\label{bound3}
\ddot{\alpha}(0) = \ddot{\alpha}_m(0), \qquad \ddot{\alpha}(\xi_f) = \ddot{\alpha}_m(\xi_f)
\end{eqnarray}
$a_j$s can be found using the boundary conditions stated in Eq.~(\ref{bound1}), (\ref{bound2}) and (\ref{bound3}). These boundary conditions preserves the widths at the beginning and the end of the waveguide and enables us to perform adiabatic like fast compression. The nonlinearity parameter $\gamma(\xi)$ can also be reverse engineered via Eq.~(\ref{alpha}) and thereby reverse engineered $\alpha(\xi)$. Note that this should also match with the initial and the final value of $\gamma(\xi)$ regardless of the propagation distance.
\begin{figure}
\onefigure[height=6cm,width=8cm]{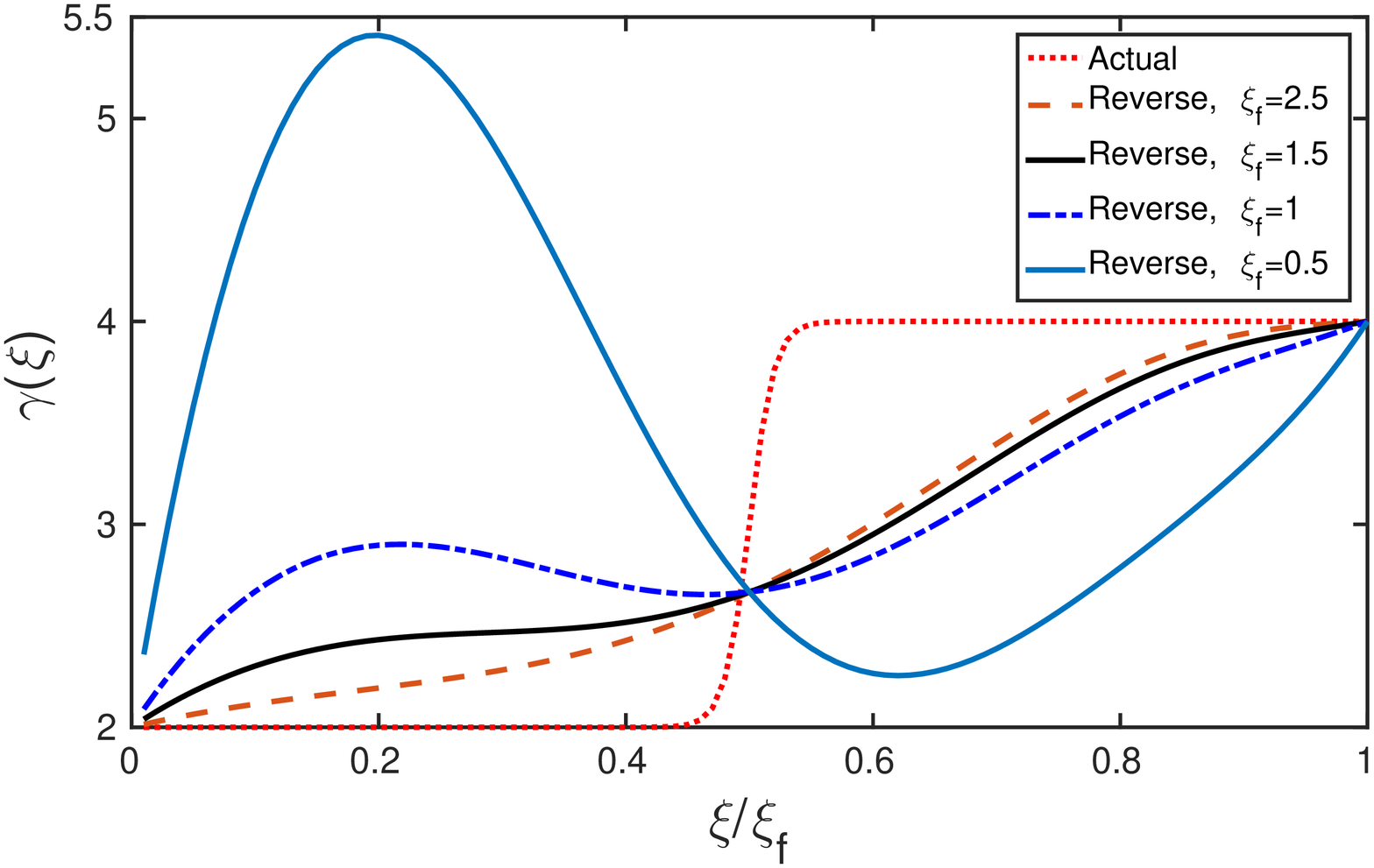}
\caption{(Color online) Comparison of reverse engineered nonlinear profile $\gamma(\xi)$ with different $\xi_f$ with $\lambda=1$ and $ \xi_f \times \delta=50$}
\label{Fig3}
\end{figure}
Fig.~(\ref{Fig2}) depicts the reverse engineered soliton pulse width and the corresponding nonlinear profile as a function of distance. It could be observed From Fig.~(\ref{Fig2}a) that the soliton pulse width does not follow the adiabatic reference, however, the initial and the final width do coincide. This clearly implies that we can obtain soliton pulse compression by using reverse engineering. Fig.~(\ref{Fig2}b) show how to engineer, with respect to the original nonlinearity given by Eq.~(\ref{NL}), the nonlinear profile of the waveguide with distance in order to achieve efficient soliton compression at a very short distance. In fact one can obtain soliton compression by this technique at an arbitrarily smaller distance, in principle. It can also be seen from Fig.~(\ref{Fig2}b) that the inverse engineered profile of nonlinearity is smoother compared to that given in Eq.~(\ref{NL}). It shows less switching rate which may be easier to design for practical purposes. 
\begin{figure}
\onefigure[height=10cm,width=8cm]{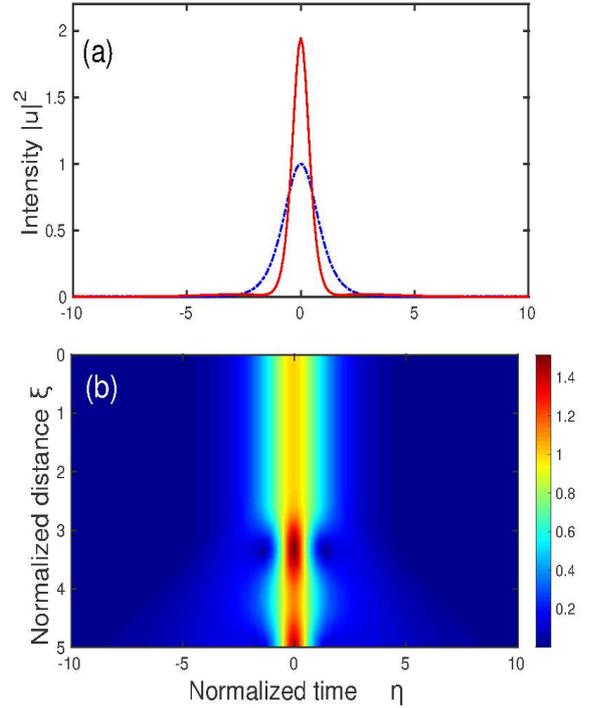}
\caption{(Color online) (a) Soliton intensity at the input (dotted blue) and the output (solid red) end. (b) Contour plot for spatio-temporal evolution of soliton intensity.}
\label{Fig4}
\end{figure}
It should be noted that, theoretically and in principle, one could compress a soliton pulse at an arbitrarily small length. However, it may not be realistic as elucidated in Fig.~(\ref{Fig3}), which depicts the nonlinear profiles for various lengths of the fiber. It is easy to observe that as one decreases the length of the fiber, the corresponding reverse-engineered nonlinear profile may no longer remains smooth enough for practical implementation. As an estimate, we may take $\xi_f=1$ or $z_f=L_D$ as the limiting value. If one takes the fiber length $z_f<L_D$, the peak value of $\gamma(\xi)$ becomes arbitrarily large and the profile shows oscillatory behaviour.

In order to illustrate the STA compression in Fig.~(\ref{Fig4}a) we plot the initial and the final soliton pulse profile. On the other hand, Fig.~(\ref{Fig4}b) depicts the contour plot for spatio-temporal evolution of soliton intensity. It can be seen that soliton propagation through the STA engineered waveguide is fairly stable. Finally, to have an idea about the utility of the proposed scheme, let us consider a silica optical fiber with $\beta_2= -20 ps^2/km $ \cite{b.b8}. Some quick calculations based on the results as illustrated above, we find that the temporal soliton pulse could be compressed from its initial pulse width say, $10 ps$ to a final pulse width $5 ps$, within a length of just $5 km$ only. This may be considered as a significant improvement over the scheme proposed by Anderson {\it et al.} \cite{b.p37.2} and, experimentally verified by by K. Bertilsson {\it et al.} \cite{b.p37.3}, where adiabatic soliton compression was achieved by engineering the dispersion but keeping nonlinear coefficient constant. K. Bertilsson {\it et al.} used a $40 km$ long fiber to compress a soliton with initial pulse width of $11 ps$ by a factor of $2.4$.One can achieve much higher compression if the nonlinearity profile is designed with a larger $\lambda$ value. We anticipate that the proposed scheme could be used for generation of ultrashort soliton pulses apart from numerous possible pulse-compression related applications.

\section{Conclusions}
In conclusion, using inverse engineering, we propose a method for temporal soliton compression in a nonlinear waveguide. We show that soliton compression could be achieved at an arbitrarily small distance with this scheme. This could possibly be exploited for various short-distance communication related applications of temporal soliton and may be even in devices.

\acknowledgments
K. P. would like to gratefully acknowledge the research fellowship provided by MHRD, Govt. of India.

\end{document}